**Unifying ultrafast demagnetization and intrinsic Gilbert damping in Co/Ni bilayers with electronic relaxation near the Fermi surface**

Wei Zhang, Wei He*, Xiang-Qun Zhang, and Zhao-Hua Cheng*

*State Key Laboratory of Magnetism and Beijing National Laboratory for Condensed Matter Physics, Institute of Physics, Chinese Academy of Sciences, Beijing 100190, P. R. China*

Jiao Teng

*Department of Materials Physics and Chemistry, University of Science and Technology Beijing, Beijing 100083, P. R. China*

Manfred Fähnle

*Max Planck Institute for Intelligent Systems, Heisenbergstraße 3, 70569 Stuttgart, Germany*

Abstract

The ability to controllably manipulate the laser-induced ultrafast magnetic dynamics is a prerequisite for future high speed spintronic devices. The optimization of devices requires the controllability of the ultrafast demagnetization time, $\tau_M$, and intrinsic Gilbert damping, $\alpha_{intr}$. In previous attempts to establish the relationship between $\tau_M$ and $\alpha_{intr}$, the rare-earth doping of a permalloy film with two different demagnetization mechanism is not a suitable candidate. Here, we choose Co/Ni bilayers to investigate the relations between $\tau_M$ and $\alpha_{intr}$ by means of time-resolved magneto-optical Kerr effect (TRMOKE) via adjusting the thickness of the Ni layers, and obtain an approximately proportional relation between these two parameters.

The remarkable agreement between TRMOKE experiment and the prediction of breathing Fermi-surface model confirms that a large Elliott-Yafet spin-mixing parameter $b^2$ is relevant to the strong spin-orbital coupling at the Co/Ni interface. More importantly, a proportional relation between $\tau_M$ and $\alpha_{intr}$ in such metallic films or heterostructures with electronic relaxation near Fermi surface suggests the local spin-flip scattering domains the mechanism of ultrafast demagnetization, otherwise the spin-current mechanism domains. It is an effective method to distinguish the dominant contributions to ultrafast magnetic quenching in metallic heterostructures by investigating both the ultrafast demagnetization time and Gilbert damping simultaneously. Our work can open a novel avenue to manipulate the magnitude and efficiency of Terahertz emission in metallic heterostructures such as the perpendicular magnetic anisotropic Ta/Pt/Co/Ni/Pt/Ta multilayers, and then it has an immediate implication of the design of high frequency spintronic devices.

*Correspondence and requests for materials should be addressed to Z.H.C (zhcheng@iphy.ac.cn) or W.H. (hewei@iphy.ac.cn)

Since the pioneering work on ultrafast demagnetization of Ni thin film after femtosecond laser irradiation was demonstrated in 1996 by Beaurepaire et al[1], the quest for ultrafast modification of the magnetic moments has triggered a new field of research: Femtomagnetism. It leads to the dawn of a new ear for breaking the ultimate physical limit for the speed of magnetic switching and manipulation, which are relevant to current and future information storage. In the past two decades, the ultrafast dynamics in hundreds of femtoseconds have been probed with the femtosecond laser pulse using magneto-optical Kerr[1] or Faraday effect[2], or other time-resolved techniques such as the high-harmonic generation (HHG) of extreme ultraviolet(XUV) radiation[3], magnetic circular dichroism[4], or spin resolved two-photo photoemission[5].

Nevertheless, the microscopic mechanism underlying ultrafast quenching of magnetization remains elusive. Various mechanisms including electron-phonon mediated spin-flip scattering[6-9], electron-electron scattering[10,11], electron-magnon scattering[12,13], direct angular momentum transfer from photon to electron mediated by spin-orbit coupling[14,15], coherent interaction among spins electrons and photons[16], were proposed to explain the ultrafast spin dynamics. In addition, since Malinowski[17] et al first proposed that the laser excited spin current transport could increase and speed up the magnetic quenching in metallic heterostructures, the laser-induced super-diffusive spin current was raised to play an important role in determining the ultrafast demagnetization in metallic films or heterostructures[18-22]. However, the recent demonstration[23] shows that the unpolarized hot electrons transport can demagnetize a ferromagnet, indicating the local spin angular momentum dissipation is unavoidable even when super-diffusive spin transport domains in the metallic heterostructures. Moreover, even in the similar samples, the local spin-flip scattering and nonlocal spin transport mechanism were proposed respectively by different experimental tools[19, 24] to explain the ultrafast demagnetization. It is harmful for clarifying the underlying ultrafast demagnetization mechanism in such metallic heterostructures. Therefore, an effective method to distinguish the two dominant contributions to ultrafast demagnetization in metallic heterostructures is highly desirable[19,23,24]. Here, we propose that investigating both the ultrafast demagnetization time and Gilbert damping[25] simultaneously is a candidate method, although the relationship between the two parameters has never been unified successfully so far between the experiments and theoretical predictions.

An inverse relation between $\tau_M$ and $\alpha_{int\,r}$ was first derived by Koopmans et al. from a quantum-mechanical calculation on the basis of the Elliott-Yalfet (EY) spin-flip scattering model[6]. Later, the attempted experiments have ever been carried out to demonstrate the prediction in rare-earth-doped permalloys[26,27] and amorphous TbFeCo films[28]. In this case, the localized $4f$ electrons rather than itinerant $5d6s$ electrons domain most of the large magnetic moment in rare-earth elements. Because the $4f$ electrons are far from the Fermi level, their ultrafast demagnetization processes are medicated by $5d6s$ electrons after laser pulse excitation[7]. The indirect excitation leads to the so called type_II ultrafast demagnetization behavior in rare-earth elements, which is much slower than that of itinerant electrons. Therefore, it is not unexpected that the ultrafast demagnetization time $\tau_M$ of permalloys increases with the doping contents of rare-earth elements increasing. Meanwhile, it happened that the Gilbert damping constant of permalloys is also increased by doping 4*f* elements, which mainly comes from the so called "slow relaxing impurities mechanism"[29]. Therefore, by introducing the extra mechanism unavoidablely, a trivial consequence was obtained that the ultrafast demagnetization time $\tau_M$ increases as the Gilbert damping $\alpha$ increases in rare-earth-doped permalloys[26]. In hindsight, from this experiment, one can not confirm the relation between ultrafast demagnetization time $\tau_M$ and Gilbert damping $\alpha$ due to the defects of the experimental design. A genuine relation between ultrafast demagnetization time and Gilbert damping should be explored in a clean system without extra demagnetization mechanism. So far, the explicit relationship between the two parameters has never been unified successfully between the experiments and theoretical predictions. Our work in Co/Ni bilayers with the electrons relaxing at the Fermi surface can fill in the blank.

In the case of pure 3*d* itinerant electrons relaxing near the Fermi surface after the laser excitation, both ultrafast demagnetization and Gilbert damping are determined by spin-flip scattering of itinerant electrons at quasi-particles or impurities. Based on the breathing Fermi-surface model of Gilbert damping and on the EY relation for the spin-relaxation time, a proportional relation between $\tau_M$ and $\alpha_{\text{int}r}$ was derived by Fähnle et al[30,31] for the materials with conductivity-like damping. And an inverse relation was also derived which is similar with that proposed by B. Koopmans et al when the resistivity-type damping domains in the materials. Although the predicted single numerical values of $\alpha_{intr}/\tau_M$ are in good agreement with the experimental ones for Fe, Ni, or Co, for a confirmation of the explicit relation between $\tau_M$ and $\alpha_{intr}$ one has to vary the values on the two parameters systematically for one system, as we do it in our paper by changing the thickness of the films.

Co/Ni bilayers with a stack of Ta (3 nm)/Pt (2 nm)/Co (0.8 nm)/Ni ($d_{Ni}$ nm)/Pt (1 nm)/Ta (3 nm) were grown on glass substrates by DC magnetron sputtering[32, 33]. The thickness of Ni layer changes from $d_{Ni}$ = 0.4 nm to $d_{Ni}$ = 2.0 nm. Their static properties have been shown in the Part I of the Supplementary Materials[34]. Both $\tau_M$ and $\alpha_{intr}$ for Co/Ni bilayer systems have been achieved by using time-resolved magneto-optical Kerr effect (TRMOKE) technique[21, 35]. The reasons for selecting the Co/Ni bilayers are three-fold. First, Co/Ni bilayers with perpendicular magnetic anisotropy (PMA) are one of candidates for perpendicular magnetic recording (PMR) media and spintronic devices[36-39]. Second, the electrons in both Co and Ni are itinerant near the Fermi surface and they have the same order of magnitude of demagnetization time[7,10]. Without rare earth element doping in 3$d$ metals, one can exclude the possibility of an extra slow demagnetization accompanied by doping with 4$f$ rare-earth metals. Third, both $\tau_M$ and $\alpha_{intr}$ in Co/Ni bilayers can be tuned by changing the Ni thickness. Therefore, Co/Ni bilayers provide an ideal system to investigate the relation between $\tau_M$ and $\alpha_{intr}$. A nearly proportional relationship between $\tau_M$ and $\alpha_{intr}$ was evident in Co/Ni bilayers, suggesting that the conductivity-like damping[30, 31] plays a dominant role. It is distinct in physics with previous experiments[26] where the seemingly similar results have been obtained via introducing extra slow demagnetization mechanism. Moreover, we discussed the origin of Gilbert damping, analyzed its influence on the relation between $\tau_M$ and $\alpha_{\text{intr}}$ and proposed a new approach to distinguish the intrinsic spin-flip and extrinsic spin current mechanism for ultrafast demagnetization in metallic heterostructures. The finding for this unification can provide the possibility for manipulating the laser-induced ultrafast demagnetization via Gilbert damping in high frequency or ultrafast spintronic devices such as the Terahertz emitters.

Fig. 1(a) shows time-resolved MOKE signals[40] for films with various Ni layer thickness measured with an external field $H = 4000$ Oe. The quantitative values of intrinsic Gilbert damping constant[41-44] in Fig.1(b) can be obtained by eliminating the extrinsic contributions (See the Supplementary Materials [34], PartⅡ for details). It was observed that $\alpha_{\text{intr}}$ decreases with increasing Ni layer thickness. On the one hand, previous investigations[39, 45] have been reported that the large PMA origins from the strong spin-orbit coupling effect at Co/Ni interface. A thickness modification in Co/Ni bilayer can change the competition between interface and volume effect, and consequently the PMA. When we plot the intrinsic Gilbert damping constant as a function of effective anisotropy field in Fig.4 in the PartⅡ in Supplementary Material(See the Supplementary Materials [34], PartⅡ for details), a proportional relation was confirmed in our Co/Ni bilayer system, which demonstrates that spin-orbit coupling contributes to both Gilbert damping and PMA (Also, for the achievement of effective anisotropy field, please see the Supplementary Materials [34] PartⅡ for details). On the other hand, the interface between Ni and Pt maybe also modified via changing Ni layer thickness. Because the Gilbert damping increases linearly when the Ni layer becomes thinner, it seems that the spin current dissipation is involved partly. A similar trend was observed in a Pt/CoFeB/Pt system[46], in which a pure non-local spin pumping effect domains the Gilbert damping. Therefore, the total Gilbert damping equals to $\alpha = \alpha_{intr} + \alpha_{sp}$, in which $\alpha_{sp}$ represents the contributions from spin current. Due to the low spin diffusion length of Pt, the magnetization precession in Ni layer entering the Pt layer would be absorbed completely like in the system of Py/Pt and Py/Pd[47] and so on. However, we have to address that, in the case of the variation of ferromagnetic layer thickness, the amount of spin current pumped out of ferromagnet is determined entirely by the parameter of interfacial mixing conductance $G_{eff}^{mix}$ [48,49]. It is a constant value once the normal metal thickness is fixed, although the Gilbert damping in thinner magnetic layer is enhanced. Therefore, given the spin current contributes partly to the Gilbert damping at present, the spin angular momentum transferring from Ni layer to Pt layer would be the same for various Ni layer thickness.

The central strategy of our study is to establish a direct correlation between ultrafast demagnetization time and the intrinsic Gilbert damping constant. The intrinsic Gilbert damping constant was extracted from magnetization precession in hundreds of ps timescale. The laser-induced ultrafast demagnetization dynamics has been measured carefully within time delay of 2.5 ps at a step of 15 fs and low laser fluence of $1\,mJ/cm^2$ was used. Fig. 2 (a) shows the TRMOKE signals of the ultrafast demagnetization evolution after optical excitation. A rapid decrease of magnetization takes place on the sub-picosecond timescale followed by a pronounced recovery. As can be seen in this figure, the ultrafast demagnetization rate is different by changing the Ni thickness.

To identify the effect of the heat transport across the film thickness on demagnetization time, a numerical simulation[50] was carried out to demonstrate that the demagnetization time variation induced with the thicknesses ranged from 1.2 nm to 2.8 nm is so small that can be ignored (See the Supplementary Materials [34], Part III for details), although a relatively large error of $\tau_M$ could be resulted in when the sample thickness spans very large. According to the simulation results, the heat transport not only affects the rate of ultrafast magnetization loss but also the maximum magnetic quenching. So, in experiment we obtain the ultrafast demagnetization time for various samples with almost the same maximum quenching of 9% to suppress the influence of heat transport[7, 21, 51-54] as well as the non local spin current effect[17]. The temporal evolution of magnetization in sub-picosecond time scale was fitted by the analytic solution based on the phenomenological three temperature model (3TM)[1, 17]:

$$-\frac{\Delta M(t)}{M} = \left\{ \left[ \frac{A_1}{(t/\tau_0 + 1)^{0.5}} - \frac{(A_2\tau_E - A_1\tau_M)}{\tau_E - \tau_M} e^{-\frac{t}{\tau_M}} - \frac{\tau_E(A_1 - A_2)}{\tau_E - \tau_M} e^{-\frac{t}{\tau_E}} \right] \Theta(t) + A_3\delta(t) \right\} * G(t, \tau_G)$$

(1)

where $*G(t, \tau_G)$ presents the convolution product with the Gaussian laser pulse profile, whose full width at half maximum (FWHM) is $\tau_G$. A temporal stretching of the laser pulse was introduced by the excited hot electrons[55], which is the trigger for the observed ultrafast demagnetization. In the fitting procedure, the demagnetization time $\tau_M$ we cared can be influenced by the value of $\tau_G$, which is inter-dependence with $\tau_M$ within the three temperature model. As is shown in Table 1 in the Supplementary Material[34] Part IV, $\tau_G$ was fixed at 330 fs for various samples to eliminate its relevance with $\tau_M$. The time variable in eq. (5) corresponds to $t = t_{\exp} - t_0$, with $t_0$ the free fit parameter characterizing the onset of the demagnetization dynamics of the actual data trace, which is fixed as 100 fs for various samples. $\Theta(t)$ is a step function, $\delta(t)$ is the Dirac delta function and $A_1, A_2, A_3$ are the fitting constants. The two critical time parameters $\tau_M, \tau_E$ are the ultrafast demagnetization time and magnetization recovery time, respectively. The well fitted curves by 3TM are also shown as the solid lines in Fig. 3(a) from which the ultrafast demagnetization time $\tau_M$ and the magnetization recovery time $\tau_E$ were evaluated. Within 3TM model, the magnetization recovery process is affected by $\tau_E$, charactering the electron-phonon relaxation, and $\tau_0$, representing heat transport timescale through the substrates as well as demagnetization time $\tau_M$. In the fitting procedure by 3TM model, we assigned a fixed value to $\tau_E$ and $\tau_0$ varies slightly to exclude the heat transport effect through thickness. Via changing the single parameter, $\tau_M$, we can accurately reproduce the experimental results for various samples. And the heat transport across the thickness domains within 3TM model characterized by the parameter of $\tau_0$, which is shown in Table. 1 in Part IV of Supplementary Material[34] as around 2 ps. It is about three times bigger than $\tau_E$ indicating that we are not mixing the heat transport and the electron-phonon relaxation[56]. Only in this case, are both the values of $\tau_E$ and $\tau_M$ genuine. The value of $\tau_0$ indicates that the heat was transferred through the substrate in less than 3 ps in this paper, rather than what was observed by F. Busse et al[57] where the heat was trapped laterally in the Gaussian profile up to 1 ns. Therefore, the lateral heat transport effect can be ignored, and hencely the modification of precessional dynamics here. As illustrated in Fig. 2(b), it can be clearly seen that $\tau_M$ decreases with increasing $d_{Ni}$.

By replotting Fig. 1(b) and Fig. 2(b), an approximately proportional relationship between $\tau_M$ and $\alpha_{intr}$ was confirmed by our experimental results (Fig.2(c)). This relationship between $\alpha_{intr}$ and $\tau_M$ is consistent well with the theoretical prediction $\tau_M \propto \alpha_{intr}$ based on the breathing Fermi-surface model[30,31,58] for materials with conductivity-like damping contributions. On the basis of the breathing Fermi-surface model, the Elliott-Yafet spin-mixing parameter $b^2$ in Co/Ni bilayers can be estimated from the theoretical equation[30, 31] shown as the red solid line in Fig. 2(c):

$$\tau_M = \frac{M}{\gamma F_{el} p b^2} \alpha \qquad (2)$$

where the quantity contains the derivatives of the single-electron energies with respect to the orientation $e$ of the magnetization M=$Me$. $p$ is a material-specific parameter which should be close to 4. If we use $F_{el}= 1.87\times10^{-23} J$ from *ab initio* density functional electron theory calculation for *fcc* bulk Ni[31], the experimental value of Elliott-Yafet spin-mixing parameter $b^2 = 0.28$ can be estimated in Co/Ni bilayers, which is far larger than that of Co or Ni. The significant enhancement of spin-mixing parameters is related to the strong spin-orbital coupling at the Co/Ni interface since $b^2$ is proportional to $\xi^2$ in first-order perturbation theory, where $\xi$ is the coefficient of the spin-orbit coupling. A detailed *ab initio calculation* for Elliott-Yafet spin-mixing parameter in Co/Ni bilayers is highly desirable. For a derivation of eq. (2) it must be assumed that the same types of spin-flip scattering processes are relevant for the ultrafast demagnetization and for the damping. The assumption does not say anything about these detailed types. It has been shown in Ref. 9 that mere electron-phonon scatterings cannot explain the experimentally observed demagnetization quantitatively. In reality there are also contributions from electron-electron scatterings[11], electron-magnon scatterings[12] and from a combination of electron-phonon and electron-magnon scatterings[13]. Because both for demagnetization and for damping, the spin angular momentum has to be transferred from the electronic spin system to the lattice, there is no reason why different types of theses spin-flip scatterings should be relevant for the two situations. Therefore, the Elliott-Yafet relation, eq. (2) should be applicable for our system. It would not be valid if non-local spin-diffusion processes would contribute a lot to demagnetization. Examples are a superdiffusive spin current in the direction perpendicular to the film plane, or a lateral diffusion out of the spot irradiated by the laser pulse and investigated by the TRMOKE. However, we definitely found the validity of the Elliott-Yafet relation, and this shows that nonlocal spin-diffusion processes are so small that can be neglected in our experiment.

Despite this, previous demonstrations[17,19-21] show that the ultrafast spin current caused by the transport of spin-majority and spin-minority electrons in the antiparallel (AP) state of magnetic multilayers after the laser pulse accelerates the ultrafast demagnetization. Similarly, as is indicated in Fig. 1(b), with the assistance of interface between FM (Ni) and NM (Pt), the spin current induced by the flow of spin-up and spin-down electrons in opposite directions[59] may contribute partly to the Gilbert damping in Pt/Co/Ni/Pt mulitilayers. The femtosecond laser induced spin current lives very shortly which is in sub-picosecond timescale, while the duration of spin current triggered by spin precession is in the timescale of nanosecond. The difference of the duration of the spin current is just related to the timescale of the perturbation of the system. One has to note that spin currents at the femtosecond time scale gives rise to a lowering of the demanetization time[17], while spin pumping induced spin current gives rise to the enhancement of Gilbert damping and thus a lowering of the relaxation time. Therefore, when spin current contributes largely to both ultrafast demagnetization and spin precession dynamics, an inverse relationship between ultrafast demagnetization time and Gilbert damping could be expected. That is, the more spin current transferred from ferromagnetic layer to normal metal, the faster ultrafast demagnetization should be. Therefore, at present paper, to explain the experimental results the local Elliott-Yafet scattering theory suffices. And, the non-local spin current effect can be ignored, although it contributes partly to the fitted value of spin-mixing parameter $b^2$ . The discussions here inspire us to continuously clarify the various relationships between ultrafast demagnetization time and Gilbert damping coming from different microscopic mechanisms, which is helpful for understanding the underlying physics of ultrafast spin dynamics as well as the application of ultrafast spin current triggered by ultrashort laser[60, 61]. For instance, recently, the researchers are seeking for the potential candidates as the Terahertz waves emitters including the metallic heterostructures. Previous demonstrations show that the magnitude and efficiency of Terahertz signals in these multilayers are determined by Gilbert damping[60]. The investigations of the relationship between Gilbert damping and ultrafast demagnetization time will open up a new avenue to tailor the Terahertz emission.

Meanwhile, the dominant contribution to ultrafast demagnetization in metallic heterostructures, either from the localized spin-flip scattering or non-local spin transport, has been a controversial issue for a long time[23]. Here, a new approach, by establishing the relation between the demagnetization time and Gilbert damping, is proposed to distinguish the two mechanisms. The proportional relationship indicates the localized spin-flip scattering mechanism domains, otherwise the nonlocal spin current domains.

In conclusion, the fast and ultrafast dynamic properties of Ta(3 nm)/Pt(2 nm)/Co(0.8 nm)/Ni($d_{Ni}$ nm)/Pt(1 nm)/Ta(3 nm) bilayers with the electrons relaxing near the Fermi surface have been investigated by using TRMOKE pump-probe technique. An genuine proportional relationship, contrast to previous trivial consequence induced by impurities mechanism, between ultrafast demagnetization time and Gilbert damping constant is confirmed from experimental results. The estimated value of spin-mixing parameter on the basis of breathing Fermi-surface model is far larger than that of Co or Ni, which is originated from the strong spin-orbital coupling at the interface. More importantly, distinguishing the dominant mechanism underlying ultrafast demagnetization in metallic heterostructures has been a tough task for a long time. Here, an effective method by unification of the ultrafast demagnetization time and Gilbert damping is proposed to solve this task, namely that, a proportional relation between the two parameters indicates the local spin flip scattering mechanism domains, otherwise the non local spin current effect domains.

**Acknowledgments**

This work was supported by the National Basic Research Program of China (973 program, Grant Nos. 2015CB921403 and 2016YFA0300701), the National Natural Sciences Foundation of China (51427801, 11374350, and 11274361). The authors thank Hai-Feng Du, Da-Li Sun and Qing-feng Zhan for critical reading and constructive suggestions for the manuscript. The authors are indebted to B. Koopmans and M. Haag for helpful discussions.

**Author Contributions**

Z.H.C. supervised project. Z.H.C. and W.Z conceived and designed the experiments. W.Z. and W.H. performed the polar Kerr loops and TRMOKE measurement. X.Q.Z. has some contributions to TRMOKE setup. T.J. provided the samples. M.F. helped with the interpretation of the results on the basis of the breathing Fermi-surface model. All the co-authors contributed to the analysis and discussion for the results. Z.H.C. wrote the paper with the input from all the co-authors.

**Additional information**

**Competing financial interests:** The authors declare no competing financial interests.

**Figure caption:**

**FIG. 1 Spin precession.** (a)TRMOKE signals of Co/Ni bilayers with $d_{Ni}$=0.4-2.0 nm in applied field $H = 4000$ Oe. (b) Intrinsic Gilbert damping constant as a function of $d_{Ni}$.

**FIG. 2 Ultrafast demagnetization.** (a) Ultrafast demagnetization curves with various Ni layer thickness. (b) Ultrafast demagnetization time as a function of Ni layer thickness. (c) Ultrafast demagnetization time as a function of Gilbert damping constant. The red full line indicates theoretical fitting.

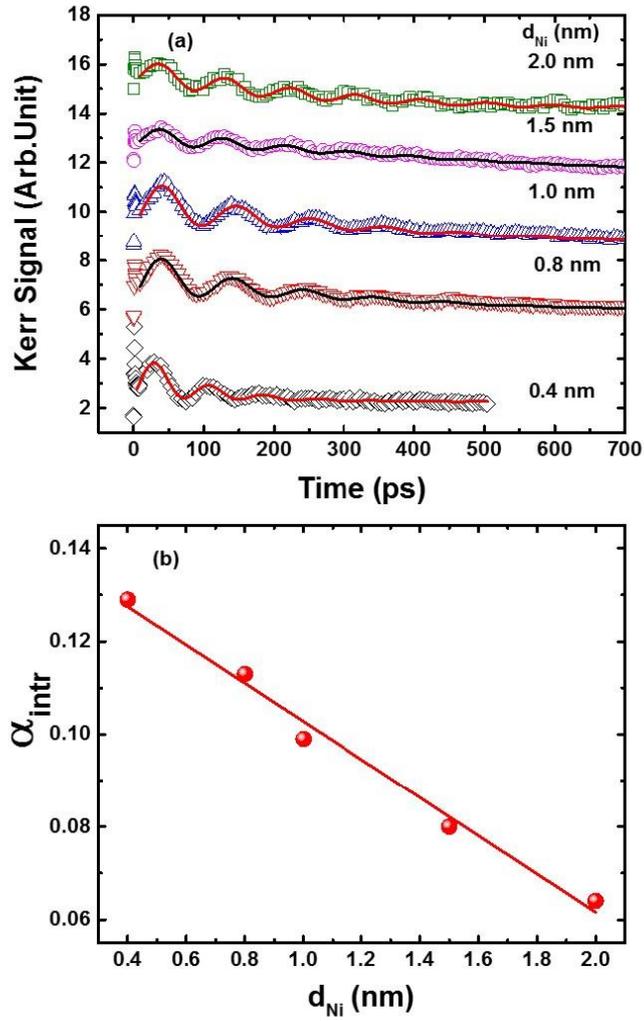

**Fig. 1 (Color Online)　Spin precession.** (a)TRMOKE signals of Co/Ni bilayers with $d_{Ni}$=0.4-2.0 nm in applied field $H$ = 4000 Oe. (b) Intrinsic Gilbert damping constant as a function of $d_{Ni}$.

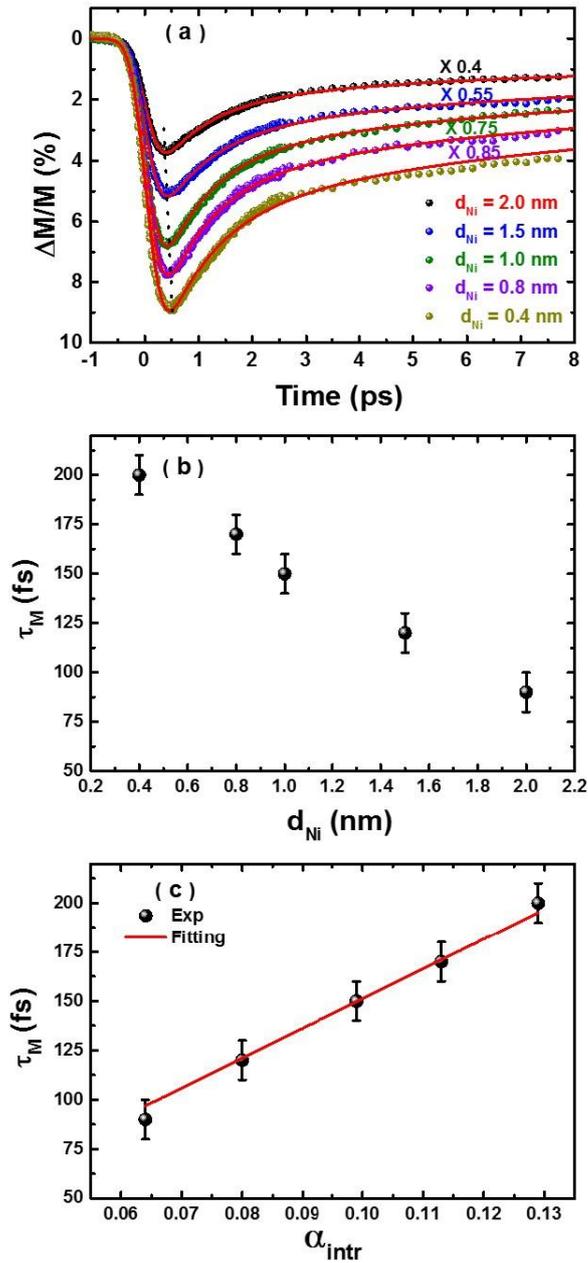

**Fig.2 (Color Online) Ultrafast demagnetization.** (a) Ultrafast demagnetization curves with various Ni layer thickness. (b) Ultrafast demagnetization time as a function of Ni layer thickness. (c) Ultrafast demagnetization time as a function of Gilbert damping constant. The red full line indicates theoretical fitting.

# Supplementary Information

**Unifying ultrafast demagnetization and intrinsic Gilbert damping in Co/Ni bilayers with electronic relaxation near the Fermi surface**

**Part I**

**The measurements of static properties for Ta (3 nm)/Pt (2 nm)/Co (0.8 nm)/Ni ($d_{Ni}$ nm)/Pt (1 nm)/Ta (3 nm).**

Fig. 1(a) shows the polar magneto-optical Kerr signal measured at room temperature with maximum applied field of 300 Oe. The static polar Kerr loops of Co/Ni bilayers were acquired using a laser diode with a wavelength of 650 nm. All samples show very square loops with a remanence ratio of about 100%, indicating the well-established perpendicular magnetization anisotropy (PMA) of the samples. The measured coercivity $H_c$ decreases with $d_{Ni}$ from 103Oe for $d_{Ni}$ = 0.4 nm to 37Oe for $d_{Ni}$ =2.0 nm (Fig. 1(b)). The decrease of coercivity implies that the PMA decreases with the thickness of Ni.

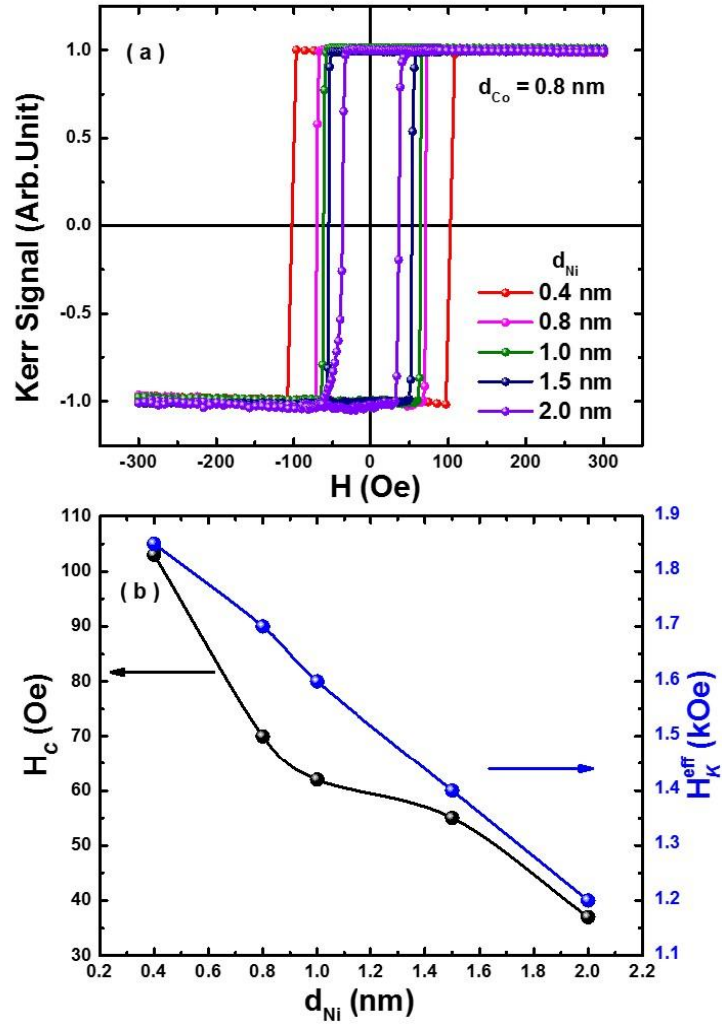

**Fig.1 Static magnetic properties of of Ta (3 nm)/Pt (2 nm)/Co (0.8 nm)/Ni (d$_{Ni}$ nm)/Pt (1 nm)/Ta (3 nm) bilayers.** (a) Polar-MOKE loops with various thickness of Ni layer d$_{Ni}$. (b) Coercivity Hc and effective anisotropy field $H_K^{eff}$ as a function of Ni layer thickness d$_{Ni}$.

## Part II

## The measurements of spin dynamics for Co/Ni bilayers in ns timescales and the analysis of extrinsic contributions to spin precession

In this part, we show the details of spin precession experiment. For example, Fig. 2(a) illustrates the scheme for laser-induced magnetization precession. The direction of applied field is fixed at $\theta_H = 80°$.

The typical time-resolved magnetization dynamics with various applied fields for Ta(3 nm)/Pt(2 nm)/Co(0.8 nm)/Ni(0.8 nm)/Pt(1 nm)/Ta(3 nm) shown in Fig. 2(b) can be best fitted by using the damped harmonic function added to an exponential-decaying background[1]:

$$\Delta M(t) = A + B\exp(-\nu t) + C\exp(-\frac{t}{\tau})\sin(2\pi f t + \varphi) \quad (1)$$

where $A$ and $B$ are the background magnitudes, and $\nu$ is the background recovery rate. $C$, $\tau$, $f$ and $\varphi$ are magnetization precession amplitude, relaxation time, frequency and phase, respectively. From the fitting curves shown in Fig. 2(b) as the solid lines, the values of precession frequency $f$ and relaxation time $\tau$ are extracted. Since the applied fields are large enough, we can obtain the Gilbert damping constant using the following relationship[2]

$$\alpha = (2\pi f \tau)^{-1} \quad (2).$$

In the case of films with a relatively low Gilbert damping[3-7] as well as thickness larger than the optical penetration depth[8], ultrafast laser may generate non-uniform spin waves and affect the relationship between demagnetization and Gilbert damping as extrinsic contributions. In order to check the contribution of non-uniform modes, we performed a fast Fourier transform shown in Fig. 2(c). Only the uniform precession mode was excited at present Co/Ni bilayers with perpendicular magnetic anisotropy.

Both $\alpha$ and $f$ as a function of $H$ are plotted in Fig.3. Since the overall damping constant consists of intrinsic damping and extrinsic damping whereby the second one arises from inhomogeneities in the sample, the Gilbert damping constant decreases monotonously to a constant value as the applied field increases (Fig. 3(a)). In the low external fields range, the inhomogeneously distributed anisotropy may lead to higher α values. Fortunately, the sufficient high field we used can suppress the extrinsic contributions to the magnetization precession, because for high fields the magnetization dynamics is mainly determined by the external field[9]. In addition, because of the interaction between femtosecond laser source and the thin films, the lateral heat distribution across the film plane has to be considered as another candidate contributions to affect the processional dynamics. As is shown by F. Busse et al[6], the heat was trapped as the Gaussian distribution across the film plane of CoFeB up to 1 ns due to the use of regenerative amplifier. It can enhance the laser power largely while the pump laser spot kept as large as around 90 μm. This facilitates the occurrence of the temperature profile, and consequently the spin-waves in the range of laser spot size. However, in the absent of regenerative amplifier at present, the laser spot is so small as less than 10 $\mu m$ [1,10] that one can excite the nonequilibrium state of the samples. And the laser fluence used here is around 1 $mJ/cm^2$, which is far weaker than that used in previous report[6]. Although smaller laser spot seems easier to trigger the nonuniform spin waves, the very low laser power we used here can suppress the influence of lateral heat distribution on the relaxation time of spin dynamics at present. Moreover, the absence of non-uniform spin wave demonstrated in Fig. 2(c) in the pump laser spot confirms that the lateral heat transport can be neglected here. In fact, it is found in the main text, within the three temperature model (3TM model) describing the ultrafast demagnetization dynamics, that the heat induced by laser pulse mainly transports along the thickness direction to substrate in less than a few picoseconds. The observation of pronounced magnetization recovery after ultrafast demagnetization can exclude the possibility of lateral heat trap.

In order to avoid the effect of extrinsic damping constant, the intrinsic damping constants were obtained by fitting the overall damping factor as the function of applied fields with the expression shown as the red line in Fig. 3(a):

$$\alpha = \alpha_{\text{int}r} + a_1 e^{-H/H_0} \qquad (3)$$

where $\alpha_{\text{int}r}$ and $a_1 e^{-H/H_0}$ are the intrinsic and extrinsic parts of the damping factor, respectively. The intrinsic part is independent of the external field or precession frequency, while the extrinsic part is field-dependent.

The experimental *f-H* relation in Fig. 3(b) can be fitted by analytic Kittel formula derived from LLG equation[2]:

$$f = \frac{\gamma}{2\pi}\sqrt{H_1 H_2} \tag{4}$$

where $H_1 = H\cos(\theta_H - \theta) + H_K^{eff}\cos^2\theta$, $H_2 = H\cos(\theta_H - \theta) + H_K^{eff}\cos 2\theta$. The equilibrium angle of magnetization $\theta$ was calculated from the relationship $\sin 2\theta = \frac{2H}{H_K^{eff}}\sin(\theta_H - \theta)$. The direction of applied field is fixed at $\theta_H = 80°$. In the above equations, $H_K^{eff}$ and $\gamma$ are the effective perpendicular magnetization anisotropy and gyromagnetic ratio, respectively, where $H_K^{eff} = \frac{2K_{eff}}{M_s}$, $\gamma = \frac{2\pi g \mu_B}{h}$. In our calculation, the Lande factor $g$ was set to 2.2 as the bulk Co value[2]. $H_K^{eff}$ is the only adjustable parameter. The variation of effective field with the thickness of Ni layer was also plotted in Fig. 1(b). When we plot the intrinsic Gilbert damping constant as a function of effective anisotropy field in Fig.4, a proportional relation was confirmed in our Co/Ni bilayer system, which demonstrates that spin-orbit coupling contributes to both Gilbert damping and PMA.

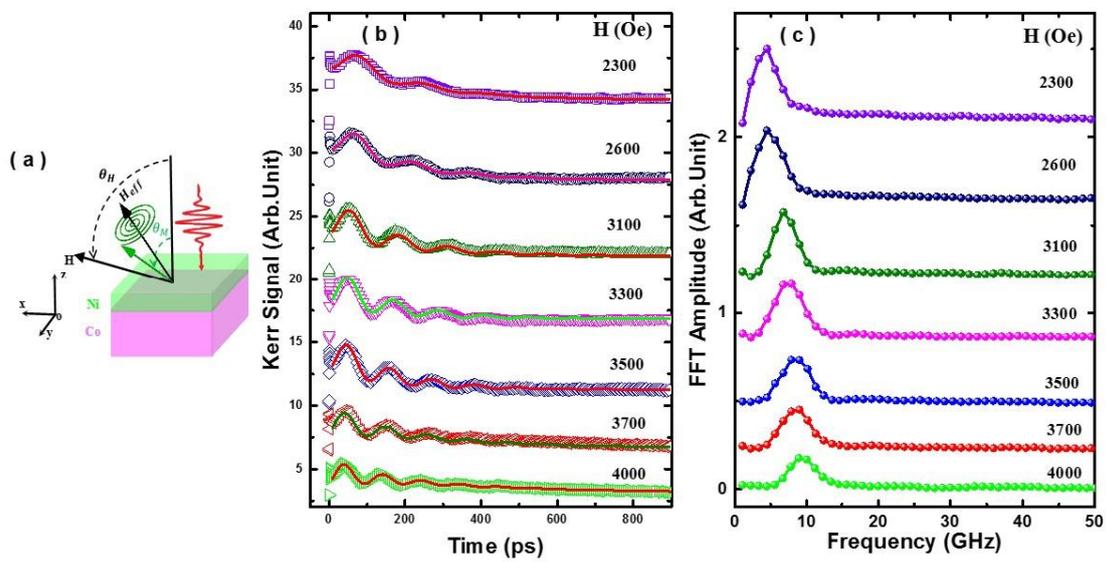

Fig. 2 (a) Scheme of TRMOKE. (b): TRMOKE signals with various applied field for Ta (3 nm)/Pt (2 nm)/Co (0.8 nm)/Ni (0.8 nm)/Pt (1 nm)/Ta (3 nm) bilayers. (c): Fast Fourier transformation signals.

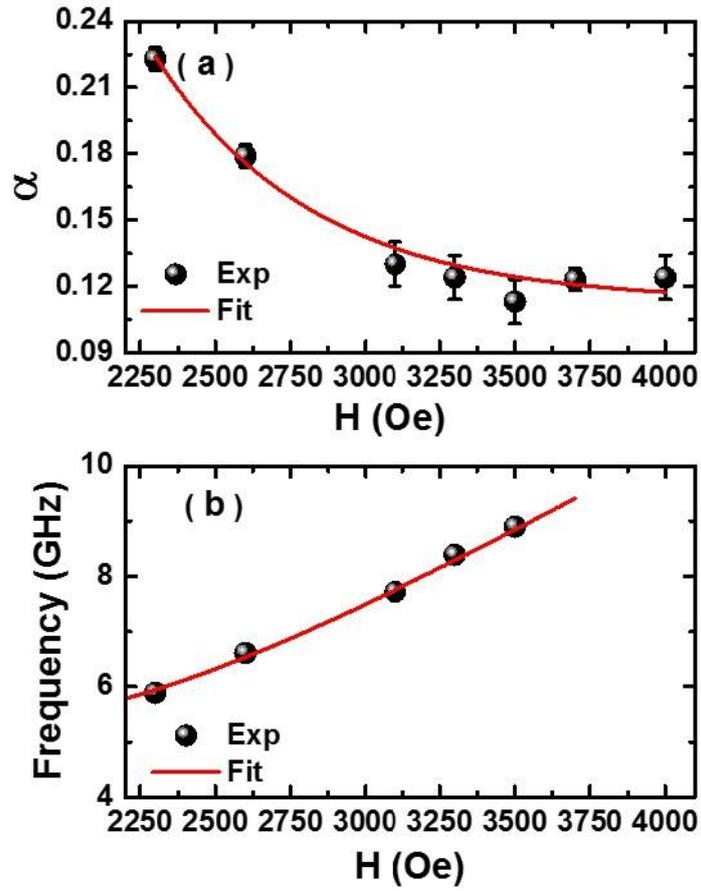

**Fig. 3 Gilbert damping and precession frequency.** Field dependence of overall damping constant (a) and precession frequency (b) of Co/Ni bilayers with $d_{Co} = 0.8nm, d_{Ni} = 0.8nm$

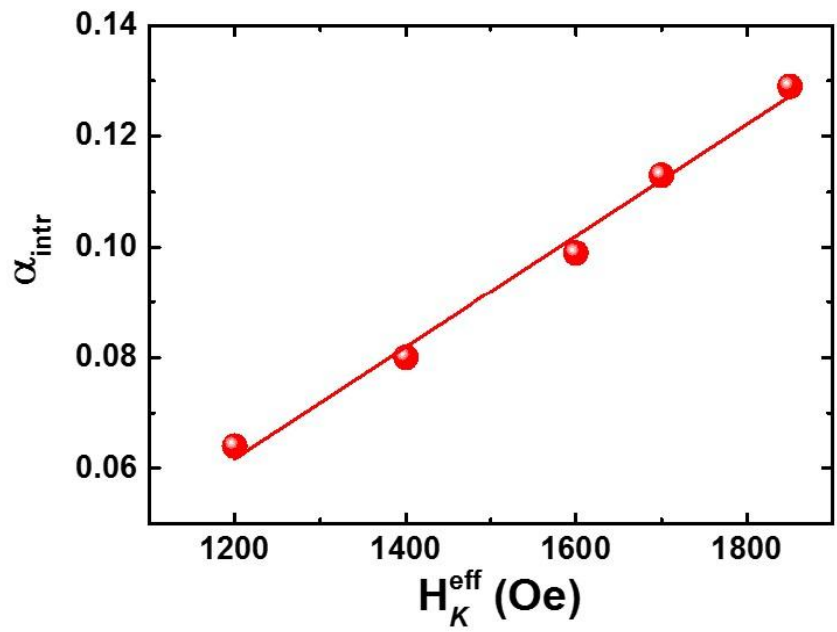

**Fig. 4   Dependence of intrinsic Gilbert damping constant on the effective anisotropy field.**

# PartⅢ

**Numerical simulation for the effect of heat transport across the film thickness on the ultrafast demagnetization time**

To estimate the evolution of heat transport profile in time, we carried out a numerical simulation based on M3TM[11] model, in which the heat transport[12] was dominated by electrons and a temperature gradient across the film thickness was introduced. It is divided in thin slabs in the direction normal to the film plane, and the slabs is 0.1 nm thick. For each slab, the evolution of the electron and phonon temperatures $T_e$ and $T_p$ are determined by a set of coupled differential equations:[13]

$$\gamma T_e(z)\frac{dT_e(z)}{dt} = \nabla_z(\kappa \nabla_z T_e(z)) + g_{ep}(T_p(z) - T_e(z)),$$

$$C_p \frac{dT_p(z)}{dt} = g_{ep}(T_e(z) - T_p(z)),$$

$$\frac{dm(z)}{dt} = Rm(z)\frac{T_p(z)}{T_c}(1 - m(z)\coth(\frac{mT_c}{T_e(z)})),$$

(5)

Where $m = \frac{M}{M_s}$, $\kappa = \kappa_0 \frac{T_e(z)}{T_p(z)}$ 4, $R = \frac{8 a_{sf} g_{ep} k_B T_c^2 V_{at} \mu_B}{\mu_{at} E_D^2}$, with $\mu_{at}$ the atomic magnetic moment in units of Bohr magneton $\mu_B$, $V_{at}$ the atomic volume, and $E_D$ is the Debye energy. $C_e$ and $C_p$ are the heat capacities of the e and p systems respectively. $\nabla_z T_e(z)$ is the electron temperature gradient normal to the film. $k_B$ is the Boltzmann constant. $k_0$ is the material dependent electronic thermal conductivity. $g_{ep}$ is the e-p coupling constant and determines the decay of the electronic temperature until equilibrium is reached[14]. $a_{sf}$ represents the spin-flip probability[11]. The equations of motion for each slab thus describe heating of the electron system by a Gaussian laser pulse, heat diffusion by electrons to neighboring slabs, e-p equilibration, and finally the evolution of the magnetization due to e-p spin-flip scattering. In the simulation, the total magneto-optical signal was obtained by the calculation of $\theta(t) \propto \int m(z,t) \exp(-\frac{z}{\lambda}) dz$.

The electronic system after the action of the laser pulse is in a strongly non-equilibrium situation. Nevertheless, one can describe the electron system by use of an electron temperature. The reason is that the laser photons excite electrons, but these excited electrons thermalize more or less instantly due to very rapid and frequent electron-electron scatterings via their Coulomb interactions. This is the assumption of the accepted Elliott-Yafet scenario which describes the effect of the laser pulse directly after the action of the laser pulse.

Fig.4(a) shows the simulated ultrafast demagnetization curves for various film thicknesses. We can clearly observe that the evolution of magnetization curves looks almost identical for various film thicknesses, indicating that the effect of heat transport on the demagnetization time can be neglected. Despite this, for the remagnetization part, a deviation from the experimental curves occurs. This is mainly because that the heat diffusion can almost be neglected during the ultrafast demagnetization timescale, but starts playing an increasing role from ps timescale onwards. The similar phenomenon was reported previously by B. Koopmans et al. Fortunately, what we should be focused on here is in the ultrafast demagnetization timescale, in which the effect of heat transport can be neglected. In fact, as is shown in Fig. 4(b), less than 10 fs variation was induced with the thicknesses ranged from 1.2 nm to 2.8 nm. The parameters used in the simulation is given in Table.1.

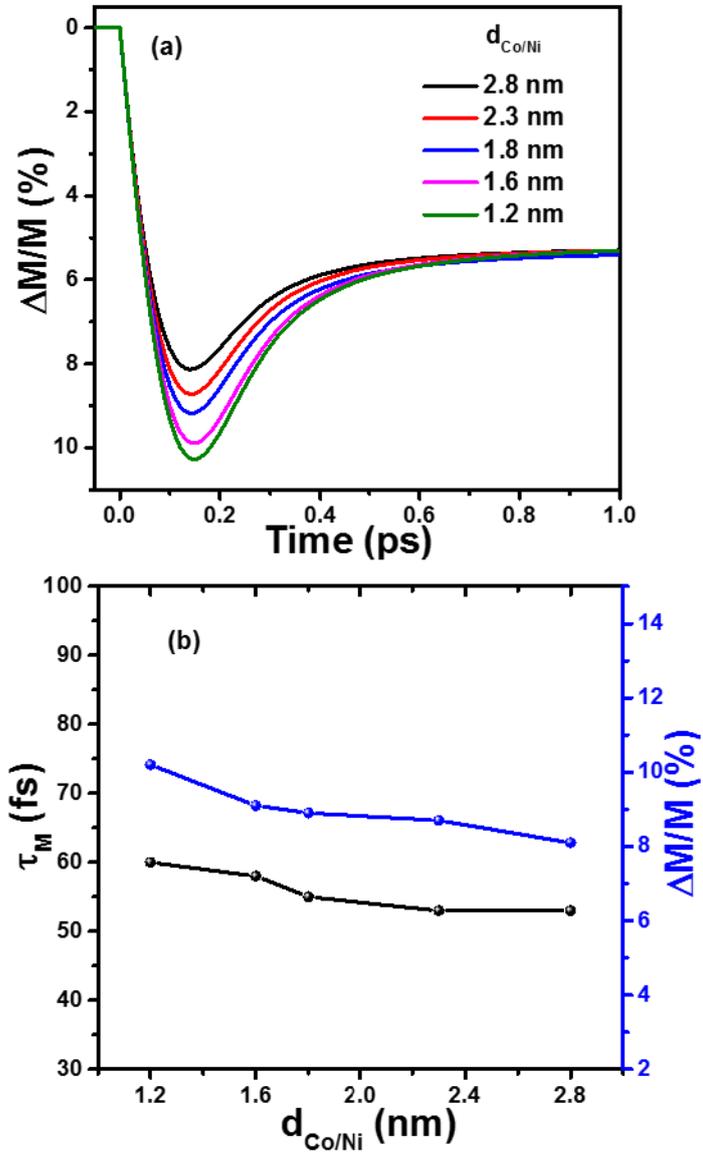

Fig. 4(a) Dependence of demagnetization as a function of delay time after pulsed laser heating at $t = 0$ (b) Maximum demagnetization and demagnetization time versus the sample thickness.

**Table 1: Parameters used in the M3TM[12,13,15].**

| Parameters | Value | Units |
|---|---|---|
| $\gamma$ | 5400 | $J/(m^3 K^2)$ |
| $C_p$ | $2.33 \times 10^6$ | $J/(m^3 K)$ |
| $g_{ep}$ | $4.05 \times 10^{18}$ | $J/(m^3 s K)$ |
| $E_D$ | 0.036 | $eV$ |
| $\mu_{at}$ | 0.62 | |
| $T_c$ | 630 | $K$ |
| $\kappa_0$ | 90.7 | $J/(smK)$ |
| $a_{sf}$ | 0.185 | |

## Part IV

**Table. 1 Values of the main fit parameters of ultrafast demagnetizations curves for various thicknesses of the samples.**

| $d_{Ni}$ (nm) | $\tau_M(fs)$ | $\tau_E(fs)$ | $\tau_0(ps)$ | $\tau_G(fs)$ | $t_0(fs)$ |
|---|---|---|---|---|---|
| **0.4** | 200 | 860 | 2.3 | 330 | 100 |
| **0.8** | 170 | 860 | 2.1 | 330 | 100 |
| **1.0** | 150 | 860 | 2.0 | 330 | 100 |
| **1.5** | 120 | 860 | 2.3 | 330 | 100 |
| **2.0** | 90 | 860 | 2.0 | 330 | 100 |